\documentclass[unpublished]{JHEP3} 
\usepackage{epsfig, subeqn}
\epsfclipon

\newcommand{\nco}{\newcommand}
\nco{\beqa}{\begin{eqnarray}} \nco{\eeqa}{\end{eqnarray}}
\nco{\lra}{\leftrightarrow}

\nco{\sss}{\scriptscriptstyle} \nco{\vpp}{\vec p^{\,2}}
\nco{\lsim}{\mbox{\raisebox{-.6ex}{~$\stackrel{<}{\sim}$~}}}
\nco{\gsim}{\mbox{\raisebox{-.6ex}{~$\stackrel{>}{\sim}$~}}}

\newcommand{\beq}[1]{\begin{eqnarray}\label{#1}}
\newcommand{\eeq}{\end{eqnarray}}

\title{\Large Yang-Type Monopoles In 5 Dimensional Curved Space-Time}

\author{Ding-fang Zeng, Wei-shui Xu \& Yi-hong Gao\\
ITP, CAS, China(Beijing)\\
E-mail: \email{dfzeng@itp.ac.cn,\ wsxu@itp.ac.cn\ \&\
gaoyh@itp.ac.cn}} \preprint{ITP-CAS 06-04} \received{\today} 

 \abstract{Motivated by Gibbons and Townsend's recent work,
 we construct Yang-Type monopoles in maximally symmetric
 space-time. We then analyze the dependence of
 horizon structure of the space-times around the
 5-dimensional monopoles on the relative strength of
 gravitations to Yang-Mills interactions. We also
 analyze the stability of the monopoles against tensor
 type perturbations on metrics.
 }

\begin{document}

 \section{Introductions}

 Recently, based on Yang's work \cite{YangMonopoles}, Gibbons and
 Townsend \cite{GibbonsTownsend} construct a class of solutions to
 Einstein-Yang-Mills system, the Yang type monopoles of gauge group
 $SO(2k+2)$. When the first version of their paper
 appears in the arXive, we are considering a related question in
 four dimensions---we are searching for numerical solutions to
 Einstein-Yang-Mills-Higgs-Kibble system \cite{LemosZanchin}, so we
 realized that their results can be generalized to asymptotically
 maximal symmetric space-times. Very quickly, Gibbons and Townsend
 revised their papers and made this generalizations. But in
 the 5-dimension case, the structure of space-time around
 the Yang-type monopoles has properties which are different from
 that of other dimensions. Besides this point, there is a free
 parameter of mass dimension, see eq(\ref{metricFunction}) in the
 following, which is unconstrained in Gibbons
 and Townsend's work. However, if the monopole solutions are to be
 stable against metric perturbations, this parameter are not totally free.
 Considering this two points, we write this paper to discuss the
 relevant questions. We also noticed an earlier work discussing
 Einstein-Yang-Mills systems in curved space-time
 \cite{BPDolan}.

 \section{General Formalisms}
 We start with the following Einstein-Yang-Mills action,
 \beq{}
 S&&\hspace{-2mm}=-\frac{1}{2\kappa^2}\int dx^D\sqrt{-g}(R+2\Lambda)
 -\frac{1}{4g_{YM}^2}\int dx^D\sqrt{-g}F^a_{mn}F^{amn}
  \label{Einstein_HKS0}
 \eeq
 where $D$ is the space-time dimension. Although our title
 contains key-words ``5-dimensional'', most of our formalisms
 apply to general dimensions. In the
 above actions, $\Lambda$ is cosmological
 constants, it can be greater than, or equal to or less than $0$,
 while $\kappa^2$ is the gravitational coupling constant. $F^a_{mn}$ fills
 adjoint representation of group $SO(D-2)$
 \beq{}
 F^a_{mn}&&\hspace{-2mm}=\partial_m A^a_n-\partial_n
 A^a_m+f^{abc}A^b_m A^c_n
 \eeq
 Relative
 to Gibbons' and Townsend's works \cite{GibbonsTownsend}, version 1,
 we include a cosmological
 constant in the Einstein-Hilbert action. In the
 second version of Gibbons's work, they have included
 cosmological constants in their basic actions.
 Because we begin to write this paper when their first
 version appears, we preserve this section in the current work as an
 adaptations for the relevant symbols.

 From the above action, we get the system's equation of motion
 \begin{subequations}
 \beq{}
 \frac{1}{\sqrt{-g}}\partial_m(\sqrt{-g}F^{mn})-i[A_m,F^{mn}]=0\label{YMeqs}\\
 R_{mn}-\frac{1}{2}(R+2\Lambda)g_{mn}+\frac{\kappa^2}{2g_{YM}^2}T_{mn}=0\label{EinsteinEqs}
 \eeq\label{EinsteinYMeqs}
 \end{subequations}
 where we have used the abbreviations $A_{m}=A^a_{m}T^a$,
 $F_{mn}=F^a_{mn}T^a$, remember that
 \beq{}
 T_{mn}=\textrm{tr}(F_{mp}F_n^p)-\frac{1}{4}g_{mn}\textrm{tr}(F_{pq}F^{pq})\\
 \textrm{[}T^a,T^b\textrm{]}=f^{abc}T^c,\ \textrm{tr}(T^aT^b)=\frac{1}{2}\delta^{ab}
 \label{Tgenerator}
 \eeq
 Based on Gibbons's and Townsend's work \cite{GibbonsTownsend}, we
 seek the solution of eqs(\ref{EinsteinYMeqs}) which has the form
 \begin{subequations}
 \beq{}
 ds^2=-f(r)dt^2+f^{-1}(r)dr^2+r^2d\Omega^2_{D-2}\label{metricAnsaltz}\\
 A_t=0,\ A_r=0,\
 A_{i}=\frac{\Sigma_{ij}n^i}{1+\sqrt{1-n^2}}\label{YMpotential}
 \eeq\label{ansaltz}
 \end{subequations}
 Where $\Sigma_{ij}s$ are constant matrix which
 span the Lie algebra $so(D-2)$, i.e.
 \beq{}
 [\Sigma_{ij},\Sigma_{kl}]=2i(\delta_{l[i}\Sigma_{j]k}-\delta_{k[i}\Sigma_{j]l})
 \label{SigmaDefinition}
 \eeq
 $\{\Sigma_{ij}\}$ can be thought as some linear transformation of
 $\{T^a\}$ defined in eq(\ref{Tgenerator}),
 $\Sigma_{ij}=\eta^a_{ij}T^a$. Note that in $SO(N)$ group, the
 number of independent generators is $\frac{1}{2}N(N-1)$, we just
 have the same number of matrices $\Sigma_{ij}$ and $T^a$.
 In eqs(\ref{YMpotential}), $n^i$ is a coordinate system
 which parametrizes the sphere $S^{D-2}$,
 they are defined through the following relations
 \beq{}
 d\Omega^2_{D-2}=(\delta_{ij}+\frac{n^in^j}{1-n^2})dn^idn^j=:h_{ij}dn^idn^j,
 \eeq
 where $n^2=\delta_{ij}n^in^j$. It should be noted that coordinate system $\{n^j\}$ covers only
 one half of the sphere $S^{D-2}$, because on the equator of this
 sphere, $n^2=1$ so the metric there is singular.

 Obviously, the only non-zero Yang-Mills field strength
 components are $F_{ij}$s. Starting from eq(\ref{YMpotential}), using
 (\ref{SigmaDefinition}) it can be shown that
 \begin{subequations}
 \beq{}
 F_{ij}&&\hspace{-2mm}=\Sigma_{ij}-\frac{2\delta_{l[i}\Sigma_{j]k}n^ln^k}{1-n^2+\sqrt{1-n^2}}\\
 F^{ij}&&\hspace{-2mm}=r^{-4}h^{ik}h^{jl}F_{kl}\nonumber\\
 &&\hspace{-2mm}=\frac{1}{r^4}(\Sigma_{ij}+\frac{2\delta_{l[i}\Sigma_{j]k}n^ln^k}{1-n^2+\sqrt{1-n^2}})
 \eeq\label{YMfieldstrength}
 \end{subequations}
 As a result of eqs(\ref{YMpotential}) and (\ref{YMfieldstrength}),
 Yang-Mills equation of motion (\ref{YMeqs})
 are satisfied automatically. So to solve the system
 governed by eq(\ref{EinsteinYMeqs}), we only need to consider its Einstein
 part. Note that, for the metric ansaltz (\ref{metricAnsaltz}),
 \beq{}
 G_{mn}:&&\hspace{-2mm}=R_{mn}-\frac{1}{2}g_{mn}(R+2\Lambda)\nonumber\\
 &&\hspace{-2mm}=\textrm{diag.}\left[
 \frac{(D-2)[(D-3)(f-1)+rf^\prime]+2\Lambda r^2}{2r^2f^{-1}},
 \right.\nonumber\\
 &&\hspace{20mm}\left.
 -\frac{(D-2)[(D-3)(f-1)+rf^\prime]+2\Lambda r^2}{2r^2f},
 \right.\nonumber\\
 &&\hspace{27mm}\left.
 -\frac{(D-3)[(D-4)(f-1)+2rf^\prime]+r^2(f^{\prime\prime}+2\Lambda)}{2r^2}g_{ij}
 \right]\label{EinsteinTensor}
 \eeq
 while
 \beq{}
 T_{mn}:&&\hspace{-2mm}=\textrm{tr}(F_{mp}F_n^p)-\frac{1}{4}g_{mn}\textrm{tr}(F_{pq}F^{pq})\nonumber\\
 &&\hspace{-2mm}=\textrm{diag.}\left[
 \frac{1}{4f^{-1}}\textrm{tr}F_{ij}F^{ij},
 \ -\frac{1}{4f}\textrm{tr}F_{ij}F^{ij},
 \ \textrm{tr}(F_{ik}F_j^{\ k})-\frac{1}{4}g_{ij}\textrm{tr}(F_{kl}F^{kl})
 \right]\nonumber\\
 &&\hspace{-2mm}=\textrm{diag.}\left[
 \frac{1}{4f^{-1}}\textrm{tr}F_{ij}F^{ij},
 \ -\frac{1}{4f}\textrm{tr}F_{ij}F^{ij},
 \ -\frac{1}{4}\frac{D-6}{D-2}\textrm{tr}(F_{kl}F^{kl})g_{ij}
 \right]\label{YMstressTensor}
 \eeq
 According to eqs(\ref{SigmaDefinition}) and
 (\ref{YMfieldstrength}), it can be verified that
 \beq{}
 \textrm{tr}(F_{ij}F^{ij})=\frac{1}{r^4}\textrm{tr}(\Sigma_{ij}\Sigma_{ij})=\frac{(D-2)(D-3)}{r^4}
 \label{traceFF}
 \eeq
 Substituting eqs(\ref{EinsteinTensor}), (\ref{YMstressTensor})
 and (\ref{traceFF}) into eq(\ref{EinsteinEqs}), take its first
 component, we obtain
 \beq{}
 \frac{(D-2)[(D-3)(f-1)+rf^\prime]+2\Lambda
 r^2}{2r^2}+\frac{\kappa^2}{2g_{YM}^2}\frac{(D-2)(D-3)}{4r^4}=0
 \label{EinsteinEq00}
 \eeq

 To solve this equation, we set
 \beq{}
 f=1-\frac{\kappa^2M(r)}{r^{D-3}}
 \eeq
 substituting it into eq(\ref{EinsteinEq00}), we will get
 \beq{}
 \frac{dM(r)}{dr}=\frac{D-3}{4g_{YM}^2}r^{D-6}+\frac{2\Lambda}{\kappa^2(D-2)}r^{D-2}
 \eeq
 so
 \beq{}
 f&&\hspace{-3mm}=\left\{
 \begin{array}{lr}
 1-\frac{\kappa^2m}{r}+\frac{\kappa^2}{4g_{YM}^2}\frac{1}{r^2}-\frac{\Lambda r^2}{3},\ D=4\\
 1-\frac{\kappa^2m}{r^2}-\frac{2\kappa^2}{4g_{YM}^2}\frac{\textrm{ln}[g_{YM}^{-2}r]}{r^2}-\frac{\Lambda r^2}{6},\ D=5\\
 1-\frac{\kappa^2m}{r^{D-3}}-\frac{(D-3)\kappa^2}{4(D-5)g_{YM}^2}\frac{1}{r^2}-\frac{2\Lambda r^2}{(D-1)(D-2)},\ D\geq6\\
 \end{array}
 \right.\label{metricFunction}
 \eeq
 where $m$ is an integration constant of mass
 dimensions. From this expression, we see that in the
 4-dimensional case, the space time is that of
 charged black-holes. Note that, the field
 produced by the charges contribute positively to gravitational
 potentials. While in 6 or greater-than-6 dimensional case, the
 field produced by the magnetic charges contribute negatively to
 gravitational potentials. Because, in the 4 dimensional
 case, our gauge group is $SO(4-2)$, an abellian group, while in
 the 6 or greater-than-6 dimensional case, the gauge group is
 $SO(D-2)$, $D\geq6$, which are non-abellian groups. We suspect the
 positive or negative property of the magnetic monopoles'
 contribution to the gravitational potentials has relevance to the
 non-abellian-ness of the underlying gauge theories. We know,
 in the higher dimensional Riessner-Nordstr\"{o}m black holes,
 the abellian $U(1)$ charge contribute to the gravitational potentials
 positively. So if we can construct magnetic monopoles out of some non-abellian gauge
 theories in 4-dimensional space-time and calculate the metrics around
 them exactly, we may find that contributions from the monopoles to
 gravitational potentials are negative instead of positive.

 From our calculations, especially the ansaltz eqs(\ref{ansaltz})
 and the final results (\ref{metricFunction}), we see that,
 adding cosmological constant in the Eistein-Hilbert action
 does not affect Yang-Mills field's configuration of the
 monopoles, i.e. in the
 asymptotically (Anti)de-Sitter space-time, Yang-Mills fields'
 configuration is completely the same as that in the
 asymptotically flat space-time. This is because, the
 Yang-Mills field of the monopoles has only non-zero
 $A_{i}$ or equivalently $A_{\theta}$, $A_{\phi}$ ... components, its
 $A_t$ and $A_r$ components are all zero. As a result, as we
 add cosmological constants in the Einstein-Hilbert action,
 although the metric's $g_{tt}$ and $g_{rr}$ are changed, its
 $g_{\theta\theta}$, $g_{\phi\phi}$ ... components does not
 change. So Yang-Mills equations of motion (\ref{YMeqs}) does not
 change as we add cosmological constants in the Einstein-Hilbert
 action.

 \section{Five-Dimensional Yang-Type Monopoles}
 In the five dimensional case, our gauge group is $SO(3)$, i.e.
 Yang-Mills field fills the adjoint representation of $SO(3)$.
 While the metric function has the form given in the second line
 of eq(\ref{metricFunction}). From the metric function, we see
 that space-time around the magnetic monopoles has very
 interesting singular structures depending on the
 relative strength of gravitations $\kappa$ to
 Yang-Mills interactions $g_{YM}$.

 \begin{figure}[h]
 \includegraphics[clip, scale=0.5, bb=53 330 505 737]{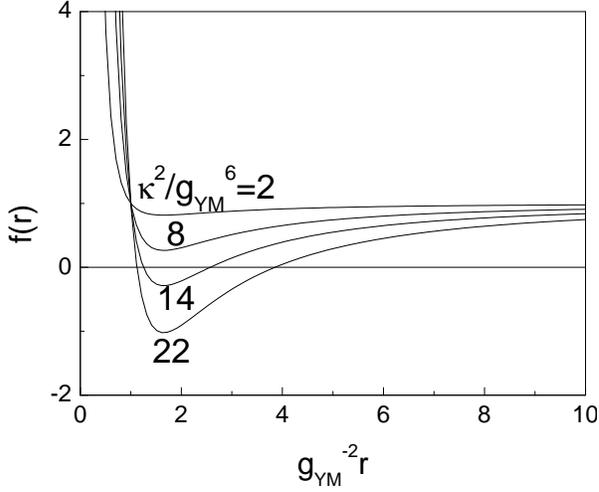}
 \caption{
 The metric function of 5-dimensional pure Yang-Type magnetic monopoles for
 five specific values of $\kappa^2/g_{YM}^6$. }
 \label{FIGmonoD5}
 \end{figure}
 First, for simplicity, we set the mass dimensional constant
 $m$ and the cosmological constant $\Lambda$ in
 eqs(\ref{metricFunction}) both to zero. In this case, we write
 $f(r)$ as
 \beq{}
 f(r)=1-\frac{\kappa^2}{2g_{YM}^6}\frac{\textrm{ln}[g_{YM}^{-2}r]}{g_{YM}^{-4}r^2}
 \label{metricFunction1}
 \eeq
 It can be easily find that
 \beq{}
 &&\hspace{-2mm}\textrm{as\ }\frac{\kappa^2}{g_{YM}^6}=4e,\ e=2.71828...\nonumber\\
 &&\hspace{-2mm}f(r)\textrm{\ has\ only\ one\ zero\ point\ }
 r=g_{YM}^2e^{\frac{1}{2}}
 \eeq
 So, as $\kappa^2/g_{YM}^6$ increases from zero to $4e$,
 the function $f$ varies from
 having no to having one zero point. The
 corresponding space-time varies from having
 no to having one horizons. Then as $\kappa^2/g_{YM}^6$ increases
 from $4e$ to infinite, the space-time varies from having one to
 having two horizons. We displayed in Figure \ref{FIGmonoD5}
 $f(r)$'s dependence on $r$ for some specific $\kappa^2/g_{YM}^6$
 values. As long as $\kappa^2/g_{YM}^6\neq0$, the point $r=0$ is
 a time-like singular point.

\begin{figure}[h]
 \includegraphics[clip, scale=0.5, bb=58 363 505 737]{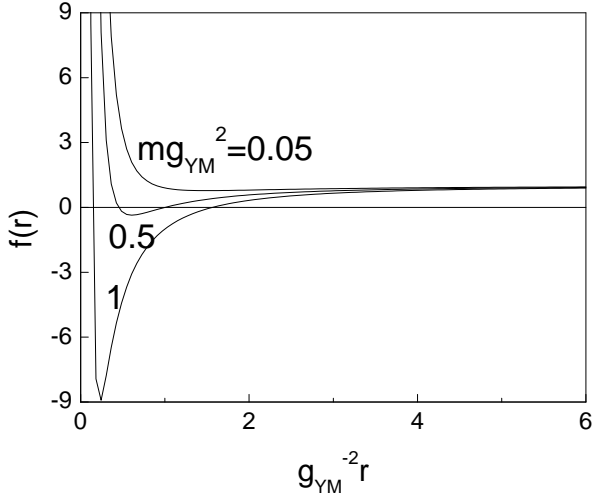}
 \caption{
 The metric function of 5-dimensional non-pure Yang-Type magnetic monopoles
 of $\kappa^2/g_{YM}^6=2$ but with three different values of $mg_{YM}^2$ }
 \label{FIGmonoD5meff}
 \end{figure}

 Second, we consider the mass parameter $m$'s effects on the
 singular structure of space-time around the monopoles. In this
 case, we write the metric function $f(r)$ as
 \beq{}
 f(r)&&\hspace{-2mm}=1-\frac{\kappa^2}{2g_{YM}^6}\frac{2mg_{YM}^2+\textrm{ln}[g_{YM}^{-2}r]}{g_{YM}^{-4}r^2}\\
 &&\hspace{-2mm}=1-\frac{\kappa^2\cdot e^{4mg_{YM}^2}}{2g_{YM}^6}
 \frac{\textrm{ln}[e^{2mg_{YM}^2}\cdot g_{YM}^{-2}r]}{e^{4mg_{YM}^2}\cdot g_{YM}^{-4}r^2}
 \label{metricFunction2}
 \eeq
 so
 \beq{}
 &&\hspace{-2mm}\textrm{as\ }\frac{\kappa^2\cdot e^{4mg_{YM}^2}}{g_{YM}^6}=4e,\ e=2.71828...\nonumber\\
 &&\hspace{-2mm}f(r)\textrm{\ has\ only\ one\ zero\ point\ }
 r=e^{-2mg_{YM}^2}\cdot g_{YM}^2e^{\frac{1}{2}}
 \eeq
 As a result, if $m>0$, we will find that relative to the zero $m$ case,
 the minimum point of the function
 $f(r)$ moves to the bottom-left direction; while if $m<0$, the
 minimum moves to the top-right directions. The larger is $|m|$,
 the more the moving amount. We displayed in Figure \ref{FIGmonoD5meff}
 $f(r)$'s dependence on $r$ for three different values of $mg_{YM}^2$
 and fixed $\kappa^2/g_{YM}^2=2$. From the figure, we can see that
 if for some too less value of
 $\kappa^2/g_{YM}^6$ and $m=0$, the space time has no horizons,
 but as long as we change $m$ to large enough values, we will find
 horizons, behind which there is another horizon or a time-like
 singularity. Contrarily, if for some very large value of
 $\kappa^2/g_{YM}^2$ and $m=0$, the space-time has two horizons, but
 as long as we change $m$ to appropriate values $m<0$, we will find
 that the horizons become degenerate and even disappears at all.

 Third, let us consider the effects of cosmological constants on
 the singular structure of space-time. For this purpose, we write
 \beq{}
 f(r)&&\hspace{-2mm}=1-\frac{\kappa^2\cdot e^{4mg_{YM}^2}}{2g_{YM}^6}
 \frac{\textrm{ln}[e^{2mg_{YM}^2}\cdot g_{YM}^{-2}r]}{e^{4mg_{YM}^2}\cdot g_{YM}^{-4}r^2}
 -\frac{\Lambda}{6}r^2
 \label{metricFunction3}
 \eeq
 So, if the
 cosmological constants are small enough, concretely
 $|\Lambda|<<e^{4mg_{YM}^2}\cdot g_{YM}^{-4}e^{-1}$, then except
 the asymptotical behavior, the singular structure of space-time
 around the monopoles keeps almost the same as that of
 eq(\ref{metricFunction2}). On the contrary,  if
 $|\Lambda|>>e^{4mg_{YM}^2}\cdot g_{YM}^{-4}e^{-1}$, then the
 space time will only have cosmic horizon determined by $\Lambda>0$,
 the Yang-Mills monopoles introduces no horizons to the space-time
 around. In the region, $|\Lambda|$ is
 comparable with $e^{4mg_{YM}^2}\cdot g_{YM}^{-4}e^{-1}$, the
 singular structure of space time will depend on the three
 parameters $\kappa^2g_{YM}^{-6}$, $mg_{YM}^2$ and $\Lambda
 g_{YM}^4$ in more complicated patterns.

 Finally, let us show that the near horizon
 geometry of the above first two cases is that of $AdS_2\times S^3$
 if the parameters of the system are chosen so that the two horizons
 degenerate. For this
 purpose we write the metric function $f(r)$ as
 \beq{}
 f(r)&&\hspace{-2mm}=1-2e\frac{\textrm{ln}r}{r^2}
 \eeq
 where we have translated all the relevant quantities dimensionless.
 Note that in this paper, $e=2.71828...$ is the base of natural logrizm,
 which has nothing to do with electro-magnetic coupling constant.
 Obviously, $f(r)|_{r=e}=0$, $f^\prime(r)|_{r=e}=0$, $f^{\prime\prime}(r)=4e^{-1}$,
 so if we let expand $f(r)$ around $r=e$, we have
 \beq{}
 f(r)=4e^{-1}(r-e)^2=:x^2
 \eeq
 as a result
 \beq{}
 ds^2=-x^2dt^2+x^{-2}dx^2+e^2d\Omega_3^2
 \eeq
 which has the standard form of $AdS_2\times S^3$.

 \section{Stability of The Monopoles Against Metric Perturbations of Tensor Type }

 In this section we study the stability of the Yang type
 monopoles under the perturbation on space-time metrics. The study
 of perturbations on the Yang-Mills field
 configurations are more complicated than that space-time metrics, we
 leave them for future works. For simplicity, we will
 focus on metric perturbations of tensor types.

 So, start with
 \beq{}
 ds^2&&\hspace{-2mm}=-f(r)dt^2+f^{-1}(r)dr^2+r^2d\Omega^2_{D-2}\nonumber\\
 f&&\hspace{-3mm}=\left\{
 \begin{array}{lr}
 1-\frac{\kappa^2m}{r}+\frac{\kappa^2}{4g_{YM}^2}\frac{1}{r^2}-\frac{\Lambda r^2}{3},\ D=4\\
 1-\frac{\kappa^2m}{r^2}-\frac{2\kappa^2}{4g_{YM}^2}\frac{\textrm{ln}[g_{YM}^{-2}r]}{r^2}-\frac{\Lambda r^2}{6},\ D=5\\
 1-\frac{\kappa^2m}{r^{D-3}}-\frac{(D-3)\kappa^2}{4(D-5)g_{YM}^2}\frac{1}{r^2}-\frac{2\Lambda r^2}{(D-1)(D-2)},\ D\geq6\\
 \end{array}
 \right.
 \eeq
 we write the tensor type perturbations of metric as
 \beq{}
 \delta g_{ab}=0,\ \delta g_{ai}=0,\ \delta g_{ij}=r^2HT_{ij}
 \label{tensorPerturbation}
 \eeq
 We will use index $a$, $b$ to denote $t$, $r$ coordinate, while
 $i$, $j$ to denote $\theta$, $\phi$ etc, $\hat{D}_i$ is the covariant
 grad operators on the transverse sphere, $D_a$ is the covariant grad
 operators in the $\{t,r\}$ space, \cite{IshibashiKodama}.
 In the above ansaltz, $T_{ij}$ is tensor harmonic function on the transverse
 sphere $S^{D-2}$ and satisfy
 \beq{}
 [\hat{D}_j\hat{D}^j+l(l+D-3)]T_{ij}=0
 \eeq
 where $l$ is the angular quantum number which takes integer values.
 Under the perturbations
 eq(\ref{tensorPerturbation}), the first order perturbed Einstein
 Equations leads to
 \beq{}
 \{D^aD_aH+(D-2)r^{-1}fH^\prime-l(l+D-3)r^{-2}H\}=0
 \label{sturmLivioul1}
 \eeq
 Let $H=\sum_\omega e^{-i\omega t}r^{-\frac{D-2}{2}}\phi(\omega,r)$,
 eq(\ref{sturmLivioul1}) becomes
 \beq{}
 \partial_r(f\partial_r\phi)
 +[\omega^2f^{-1}-(\frac{l(l+D-2)}{r^{2}}+\frac{(D-2)(D-4)f}{4r^2}
 +\frac{(D-2)f^\prime}{2r})]\phi=0
 \label{SchrodingerLike1}
 \eeq
 Introducing tortoise coordinate $dr_\star=f^{-1}dr$, the above
 equation further simplifies to
 \beq{}
 &&\hspace{-5mm}\partial_\star\partial_\star\phi+[\omega^2-V(r_{\star})]\phi=0\nonumber\\
 &&\hspace{-5mm}V(r_\star)=(\frac{l(l+D-3)}{r^2}+\frac{(D-2)(D-4)f}{4r^2}+\frac{D-2}{2r}f^\prime)f
 \label{asympBasicEq}
 \eeq

 In the stability analysis of black
 holes in higher dimensions \cite{IshibashiKodama}, we know that if the operator
 \beq{}
 A=-\partial_\star\partial_\star+V(r_\star)
 \label{stableOperatoreA}
 \eeq
 appearing in eq(\ref{asympBasicEq}) is positive and
 symmetric with respect to the inner product
 \beq{}
 <\phi_1,\phi_2>=\int_{-\infty}^{\infty}
 dr_\star\bar{\phi}_1(r_\star)\phi_2(r_\star),
 \eeq
 then the space-time
 configuration is stable under tensor perturbations. A
 statement which is more simple in operations about
 the stability of black holes against perturbations is, as long as the
 potential appears in eq(\ref{stableOperatoreA}) is positive (or although
 contain negative part but the minimum is finite) in
 the range $r_{horizon}<r<\infty$, the space-time around the black
 hole is stable, otherwise it is unstable.

 Now we borrow this result into our analysis of Yang type monopoles.
 In the case of 4 dimensions, if the mass parameter $m\geq(\kappa
 g_{YM})^{-1}$, the space time around the monopole is a standard
 Riessner-Nordstr\"{o}m black hole, and the potential $V(r_\star)$
 is positive in the range $r_{horizon}<r<\infty$, so the space-time
 configuration around it is stable. On the contrary, if $m<(\kappa
 g_{YM})^{-1}$, the space-time around the monopoles has no
 horizon, and the potential function of eq(\ref{stableOperatoreA})
 $V(r_\star)\rightarrow-\infty$ as $r\rightarrow0$, so we conclude
 that the space-time configuration around the 4-dimensional Yang-type
 monopoles is unstable against tensor perturbations if
 the mass parameter $m<(\kappa g_{YM})^{-1}$. Considering the
 effects of cosmological constants does not alter this conclusion \cite{IshibashiKodama}.

 In 6 or higher dimensional cases, the space-time around the
 monopoles are different from that of higher dimensional
 Riessner-Nordstr\"{o}m black holes since the sign of the magnetic
 charge contributions to gravitational potentials are negative.
 Because of this, as long as the mass parameter $m>0$, the space
 time around the monopoles has non-zero horizon radius and the potential
 function of eq(\ref{stableOperatoreA})
 $V(r_\star)>0$ as $r_{horizon}<r<\infty$, so we conclude
 that the space-time configuration around the 6 or greater-than-6
 dimensional Yang-type monopoles is stable against tensor perturbations if
 the mass parameter $m>0$. If $m<0$ but $|m|$ is
 small enough so that space-time has non-zero horizon
 radius, the space-time configurations around the monopoles will
 be still be stable. It can be shown that the sufficient condition
 that space-time has non-zero horizon radius is
 \beq{}
 m>-\frac{1}{(D-5)2^{D-4}}\left[\frac{\kappa}{g_{YM}}\right]^{D-3}
 \eeq
 However, if $m<0$ but $|m|$ is so large that the
 space-time has only zero horizon radius, the potential function
 in eq(\ref{stableOperatoreA}) $V(r^\star)\rightarrow-\infty$ as
 $r\rightarrow0$, so the space-time
 configurations will be unstable against tensor perturbations.
 Considering the effects of cosmological constants does not alter
 this conclusion \cite{IshibashiKodama}.

 From the above analysis, we find that as long as the
 parameters of the system are
 chosen so that non-zero horizon radius exist, the space-time
 configurations around the monopoles will be stable. Otherwise,
 the space-time configuration will be unstable. We verified that
 this is the case in 5 dimensions either.

 \section{Summaries}

 In the first section of this paper, we generalize Gibbons and
 Townsend's construction of Yang-type monopoles to maximally
 symmetric space time and find that the asymptotical property of
 space-times does not change the configuration of the monopoles.
 In the second section of the paper, we analyzed the singular
 structure of the space-time around the five dimensional Yang-type
 monopoles. We find that, even the pure Yang-type monopoles, which
 has no explicitly introduced masses, could have non-trivial
 horizon structures as long as the relative strength of
 gravitations and Yang-Mills interactions is chosen appropriately.
 This is a phenomenon which is not mentioned by Gibbons and
 Townsend. In the final section of the paper, we analyze the
 stabilities of the Yang-Type monopoles under the tensor type
 perturbations on space-time metrics. We find that as long as the
 parameters of the system are chosen so that non-trivial horizons
 exists in the space-time around the monopoles, the space-time
 configuration will be stable against the perturbations.

 We wish our works will be relevant in the study of AMNV's
 weak gravitation conjectures using Ads/CFT. Because,
 as we have seen in the five dimensional case, the relative strength of gravitations
 and Yang-Mills interactions directly determines the singular
 structure of the space-time. While ANMV's conjectures are
 just about the relative strength of gravitations and some
 $U(1)$ gauge interactions. We know that the magnetic monopoles
 and $U(1)$ gauge interactions
 usually occurs accompanying the spontaneous break down of some larger
 non-abellian gauge symmetric theories.

\end{document}